\documentclass[12pt]{article}
\usepackage[dvips]{color}
\usepackage{epsfig}
\usepackage{amsmath}
\usepackage{graphicx}

\title{\bf Charged accelerating AdS black hole of $f(R)$ gravity and the Joule-Thomson expansion}
\author{ \small M. Rostami
\footnote{M.rostami@iau-tnb.ac.ir}~$^1$, J. Sadeghi \footnote{pouriya@ipm.ir}~$^2$, S. Miraboutalebi \footnote{S$_{-}$mirabotalebi@iau-tnb.ac.ir }~$^1$,  A. A. Masoudi \footnote{masoudi@alzahra.ac.ir}~$^{1,3}$,  { B. Pourhassan
\footnote{b.pourhassan@du.ac.ir}~$^4$,} \\ \\
$^1${\small Department of Physics, North Tehran Branch, Islamic Azad University, Tehran, Iran.}\\
$^2${\small Department of Physics, Faculty of Basic Sciences, University of Mazandaran, Babolsar, Iran.}\\
$^3${\small Department of Physics, Alzahra University, Tehran, Iran.}\\
$^{4}${\small School of Physics, Damghan University, Damghan, 3671641167, Iran.}}
\begin{document}\maketitle

\begin{abstract}
In this paper, the thermodynamical properties and the phase transitions of the charged accelerating anti-de Sitter (AdS) black holes are investigated in the framework of the $f(R)$ gravity. By studying the conditions for the phase transitions, it has been shown that the $P-V$ criticality and the van der Waals like phase transitions can be achieved for $ T \approx T_{c} $.  The Joule-Thomson expansion effects are also examined for the charged accelerating AdS black holes of the $f(R)$ gravity.
Here, we derive the inversion temperatures as well as the inversion curves. Then, we determine the position of the reverse point for different values of mass $M$  and parameter $b$ for the corresponding black hole. At this point, the Joule-Thompson coefficient is zero. So, in such case,  we can say that such point is very important for the finding of cooling - heating regions. Finally, we calculate the ratio  of minimum inversion temperature and critical  temperature for  such black hole.\\\\
{\bf Keywords:} Phase transition; Thermodynamic; Black holes; $f(R)$ gravity.
\end{abstract}

\section{Introduction}
The investigation of the black holes thermodynamics can improve our knowledge about quantum gravity as the Hawking radiation has firstly sparked our curiosity about it and its consequences \cite{1, 2, 3, 4, 5, 6}. Hawking and Bekenstein have firstly demonstrated a law of  black holes mechanics which with suitable definitions of temperature, entropy, and energy leads to a well-known thermodynamics law. Also, Hawking and Page have done a study on the thermal properties of  AdS black hole and showed that there exists a phase transition in the phase space between the Schwarzschild-AdS black holes and the thermal radiation \cite{7}. Recently, the relation between the variation of the cosmological constant $ \Lambda $ with the first law of the black hole thermodynamics is attended by several authors. In this case, the parameter $ \Lambda $ is explicitly appeared in the pressure and hence in its conjugate variable namely, the volume \cite{8}. In fact, by using the unites $ G_{N} =\hbar  = c = \kappa= 1$,  the pressure is identified by the expression,
\begin{equation} \nonumber
P = - \frac{\Lambda}{8 \pi} = \frac{3}{8 \pi} \frac{1}{\ell^{2}}~,
\end{equation}
where $ \ell$ is the length of  AdS-black hole. Also, the corresponding volume is given by,
\begin{equation} \nonumber
V = ( \frac{\partial M}{\partial P})_{S,Q,J}~,
\end{equation}
where $ M $ is the black hole mass.\\
The presence of a negative cosmological constant is significant in holography and  AdS/CFT correspondence \cite{9, 10, 11}. Indeed, the asymptotic AdS black hole space-times manages exactly to the gauge duality description with dual thermal field theory from the AdS/CFT point of view.
This corresponding theory, in fact, brings us to the interesting phenomenon known as the Hawking-Page's phase transition \cite{7}.\\
The consequent thermodynamical properties help us to understand that phase transition of charged AdS black hole has similar behavior with van der Waals fluid \cite{12, 13, 14, 15, 16, 17, 18, 19, 20}. The van der Waals behavior of black holes is attractive since it connects between black hole thermodynamics and an ordinary thermodynamical system, and this duality helps to understand about the black hole physics \cite{UP}.
This situation is not valid for every black holes, for example, in the Ref. \cite{massive1} it is found that the massive charged BTZ
black hole \cite{massive2} has not any dual van der Waals fluid, while logarithmic corrected AdS black hole in massive gravity is a holographic dual of van der
Waals fluid \cite{massive3}. Also, In the Ref. \cite{NPB} it has been found that five-dimensional singly spinning Kerr-AdS black hole behaves as van der Waals
fluid. It help us to fix some model parameter in the experimental with van der Waals
fluid and apply it to the corresponding black hole.\\
Several authors have discussed the phase transition in different black holes. As an example, the Reissner-Nordstr\"{o}m-Anti de Sitter (RN-AdS)  black hole has a first-order phase transition which is similar to the phase transition of the Van der Waals fluid \cite{22, 23, 24, 25}. In fact, the phase transition plays an important role to investigate  the thermodynamical properties of the objects in the critical point.\\
It should be noticed that there are several ways to study the phase transition. Here, we apply the heat capacity approach in different ensembles. In that case, there are two types of phase transitions. In the first type, the change of the sign and the roots of the heat capacity gives us the phase transition. But,  the second type complectly  related to the divergencies of the heat capacity (asymptotic behavior). It means that the singularity of the heat capacity show the phase transition \cite{7, 26, 27}. On the other hand, the stability and instability of the black hole can be obtained by heat capacity.
In general black hole backgrounds, heat capacity is always negative which shows that the black hole is unstable and have Hawking radiation.
But the sign of the heat capacity can be changed by presence of charge and rotation. In that case, the heat capacity is positive and the phase transition occurs \cite{29}.\\
On the other hand, the inflation theory and dark energy lead us to modified gravity of Einstein. So, in that case the $f(R)$ gravity is an example of modified Einstein$^{,} $s gravity. In general, by adding  this $f(R)$ as a higher powers of the $R$,  one can make the Ricci and Riemann tensors and  their derivatives in the Lagrangian \cite{30, 31, 32, 33, 34, 35, 36, 37, 38}. Meanwhile, considerable effort has been made to investigate the AdS black hole
thermodynamics in the background of $f(R)$ gravity with constant curvature \cite{39}.
In that case, already the phase extended of space, critical behavior of the van der Waals-like fluid and canonical ensemble for the black holes are investigated by the Refs. \cite{40, 41, 42, 43}.\\
Here, we study the thermodynamical structure for a configuration of the charged accelerating AdS black hole as well as its Joule-Thomson expansion,  in the presence of the $f(R)$ background.  Accelerating black holes are interesting backgrounds in theoretical physics. Accelerated charged rotating black holes already studied by the Ref. \cite{A0}. Accelerated charged black holes in AdS (dS) space-time, which originality constructed by the Ref. \cite{A00}, considered already to investigate gravitational and electromagnetic radiation \cite{A1} (\cite{A2}). Pair of accelerated black holes in AdS background considered by the Ref. \cite{A3}. Thermodynamics of accelerating black holes have been study by the Refs. \cite{A4, A5, A55} while holographic thermodynamics of this black hole considered by the Ref. \cite{A6}. Recently, the thermodynamic phase behavior of rotating and slowly accelerating AdS black holes investigated by the Ref. \cite{A7} and the thermodynamic behavior of charged rotating accelerating
black holes is investigated by the Ref. \cite{A77}. Also chemical variables of such black hole obtained by the Ref. \cite{A8}. While writing this paper,a charged accelerating AdS black hole solution in $f(R)$  gravity obtained and its thermodynamic behavior investigated \cite{A9}.\\
Recently, Refs. \cite{44, 45, 46}  have analyzed the Joule-Thomson expansion for AdS black holes, which yield similar results to that of the properties derived from van der Waals fluid. In general, one can say that the Joule-Thomson expansion of a gas system occurs in constant enthalpy which for a system of a black hole is defined by mass \cite{47, 48}. Joule-Thomson expansion approach is suitable for an isenthalpic system in which a thermal system is described by thermal expansion. In this expansion, the Joule-Thomson coefficient is given by $ \mu = \left( \frac{\partial T}{\partial P}\right)_{H}$,  which characterized the cooling and the heating zones. It should be noticed that, by the expansion of a thermal system with temperature $T$, the pressure always reduces and leads to a negative pressure gradient $\partial P$ \cite{44}. In this regard, one can define two different regions concerning the inverse temperature $T_{i}$ where the Joule-Thomson coefficient vanishes. When the system temperature reaches the $T_{i}$, the pressure reaches the reverse $ P_{i}$, so we define a specific point called the reverse point $( T_{i}, P_{i}) $ where cooling and heating transition occur. The upper area of the inverse curve is the cooling region, while the lower area of the inverse curve is related to the heating region.\\
All above information motivates us to arrange our paper as follows: In section 2, the charged accelerated  AdS black hole in the $f(R)$ gravity considered. In section 3, the thermodynamical properties of the considered black hole are investigated in the extended phase space. Then, by using the obtained thermodynamical properties, the critical values of such a black hole are calculated and eventually, the phase transitions and the stability of the system are examined.
In section 4, the Joule-Thomson expansion for the charged accelerated AdS black hole in the presence of the $f(R)$ gravity examined. Then, under the assumption of the constant enthalpy the inverse temperature as the inverse curves obtained. Also, the position of the reverse point is determined to respect the different values of the mass $M$  and the parameter $b$ for such a black hole. At this point, the Joule-Thompson coefficient $\mu$ is zero. This important property helps us to restrict the cooling-heating regimes by examining the sign of $\mu$. Finally, the last section is devoted to the conclusions.
\section{Charged accelerating AdS black hole in $f(R)$ gravity}
\begin{figure}[h!]
 \begin{center}$
 \begin{array}{cccc}
\includegraphics[width=75 mm]{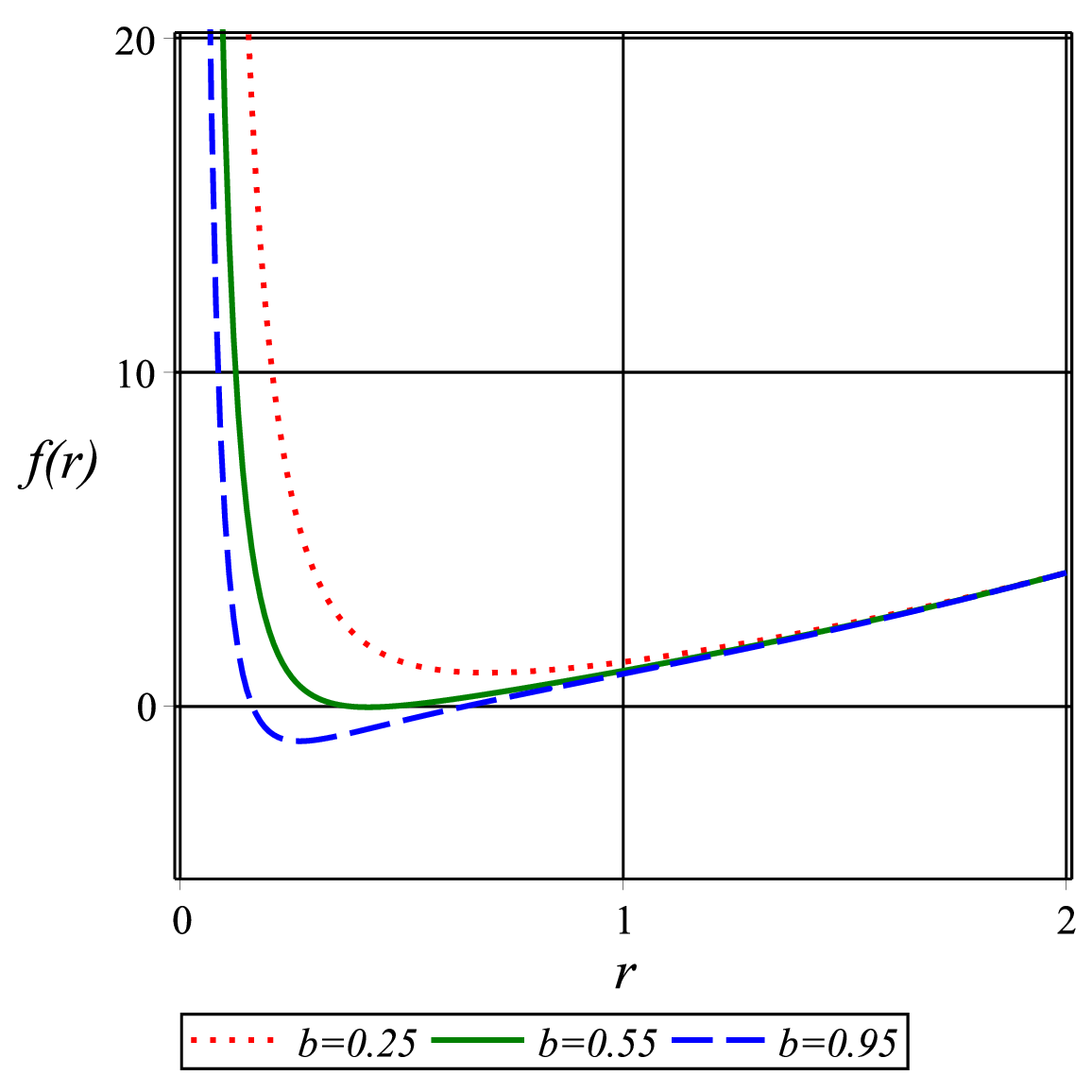}
 \end{array}$
 \end{center}
\caption{The function $f(r)$ in terms of the horizon radius with different values of $ b $, and $A=0.5$, $q = 0.4$, $R_{0}=-12$, $m = 0.6$.}
 \label{fig1}
\end{figure}

Let us consider the charged accelerating  AdS black hole in the $f(R)$ gravity. In order to study the phase transition of this black hole, its thermodynamical properties should be studied. The corresponding black hole solution is given by \cite{49,50},
\begin{equation}\label{1}
ds^{2}=\frac{1}{\Omega^{2}}\bigg[f(r)dt^{2}-\frac{dr^{2}}{f(r)}-r^{2}\bigg(\frac{d\theta^{2}}{g(\theta)}+g(\theta)\sin^{2}\theta
\frac{d\phi^{2}}{K^{2}}\bigg)\bigg],
\end{equation}
with,
\begin{eqnarray}\label{2}
f(r)&=&(1-A^{2}r^{2})(1-\frac{2m}{r}+\frac{q^{2}}{r^{2}})+\frac{r^{2}}{\ell^{2}},\nonumber\\
g(\theta)&=&1+2mA\cos\theta+q^{2}A^{2}\cos^{2}\theta,
\end{eqnarray}
where $\Omega$ is the conformal factor and is given by,
\begin{equation}\label{3}
\Omega=1+Ar \cos\theta.
\end{equation}
The parameters $m$, $q$ and $A > 0$ specify the mass, electric charge and the magnitude of the acceleration of the black hole, respectively, also  $ \ell $ is its AdS radius.  According to behavior of $g(\theta)$ at poles ($\theta_{+} = 0$ and $\theta_{-} =\pi$) and angular part of the metric, one yields to the presence of the cosmic string \cite{30, PLB}.\\
The regularity of the metric at the poles leads us to arrange the following equation,
\begin{equation}\label{4}
K_{\pm}=g(\theta_{\pm})=1 \pm 2mA + q^{2}A^{2}.
\end{equation}
Here, it should be noted that for $mA \neq 0$, the parameter $K$ can not be fixed at both poles in such a way that the
regularity exists at both poles.
%The lack of regularity of the system shows the presence of the conical singularity. For this reason,
%$K$ is chosen to regularize at one pole, leaving either a conical
%deficit or a conical excess along the other pole. Since a conical
%excess is created by a negative energy object,
It is assumed that the black hole  is regular on the north pole $(\theta = 0)$ with $K = K_{+} = 1+ 2mA + q^{2}A^{2}$ and then on the south pole ($ \theta=\pi $).  So, in that case, there is a conical deficit,
\begin{equation}\label{5}
\delta = 2\pi (1-\frac{g_{-}}{K_{+}})=\frac{8\pi mA}{1+2mA+q^{2}A^{2}},
\end{equation}
which is corresponding to a cosmic string with tension $\mu = \frac{\delta}{8\pi}$.\\
Now, we are going to review the thermodynamics of the above mentioned charged accelerating AdS black
hole. The mass of the black hole $M$, the electric charge $Q$ and the
electrostatic potential $\Phi$ are given by,
\begin{eqnarray}\label{6}
M&=&\frac{m}{K}, \nonumber\\
Q&=&\frac{1}{4\pi}\int_{\Omega=0}\ast F=\frac{q}{K}, \nonumber\\
\Phi&=&\frac{q}{r_{h}},
\end{eqnarray}
where $F$ is the electromagnetic field tensor which is related to the
gauge potential $B$ as,
\begin{equation}\label{7}
F=dB   ~~~~~~~,~~~~~~B=-\frac{q}{r}dt.
\end{equation}
The entropy of the corresponding black hole will be as,
\begin{equation}\label{8}
S=\frac{\pi r_{+}^{2}}{K(1-A^{2}r_{+}^{2})},
\end{equation}
where the Wald method \cite{wald} (instead of area law) is used to obtain it. Now, we are going to write the following action of the black hole  with $f(R)$ gravity \cite{39},
\begin{eqnarray}\label{9}
\mathcal{S} = \int_{M}{ d^4 x \sqrt{-g}\left[ r + f(R) - F_{\mu \nu} F^{\mu \nu} \right]},
\end{eqnarray}
where $f(R)$ is a function of $R$,  and  $ F_{\mu \nu} = \partial_{\mu} A_{\nu} - \partial_{\nu} A_{\mu}  $ is the electromagnetic field tensor. From the action (\ref{9}), the equations of motion for gravitational field $ g_{\mu \nu}$  and the gauge field $ A_{\mu} $ are given by,
\begin{eqnarray}\nonumber
R_{\mu \nu}[ 1 + f^{\prime}(R) ] - \frac{1}{2} g_{\mu \nu} [ R + f(R) ] + (g_{\mu \nu}\nabla^{2} - \nabla_{\mu} \nabla_{\nu} ) f^{\prime} (R)  = T_{\mu \nu},
\end{eqnarray}
\begin{eqnarray}\label{10}
\partial_{\mu}(\sqrt{-g} F^{\mu \nu}) = 0,
\end{eqnarray}
respectively.\\
For simplicity,  we consider only the case  of  constant scalar curvature $ R = R_{0} $ for the gravity.  So, the solution of equation (\ref{9}) with $  f(R) $ at constant scalar curvature $ R = R_{0} $ is given by,
\begin{eqnarray}\label{11}
f(r)=(1-A^{2} r^{2})(1-\frac{2m}{r}+\frac{q^{2}}{b r^{2}}) - \frac{ R_{0}r^{2}}{12},
\end{eqnarray}
where  $ b = [ 1+ f^{\prime} (R_{0})] > 0 $ and $ R_{0} < 0 $ corresponding to asymptotically AdS solution.\\
In Fig. \ref{fig1} we draw the $f(r)$ in terms of the radius to study horizon structure of the black hole, and see the effect of $f(R)$ gravity (parameter $b$) on the corresponding black hole. We can see three different situations according to the value of $b$. For the selected values of $m$ and $q$, there is a critical value $b_{c}$ where the black hole is extremal ($b_{c}=0.55$ corresponding to solid green line of the Fig. \ref{fig1}). In the cases of $b<b_{c}$ we see only a naked singularity, while for the cases of $b>b_{c}$ we have regular black hole with inner and outer horizons. We can also see the effect of $b$ parameter in the small $r$. It means that for the larger $r$ the function $f(r)$ does not affected by $b$. In another word, we can say that $f(R)$ gravity effect is negligible for the larger black hole. It behaves like quantum corrections due to thermal fluctuations \cite{log1} which are important for the small black holes \cite{log2}. Hence, the $f(R)$ gravity terms do not more affect the metric function when $r$ is large.
Generally,  one can say that for the small values of $b$ parameter, the metric function changes. But, the metric function does not more change when parameter $ b $ is large.\\

\begin{figure}[h!]
 \begin{center}$
 \begin{array}{cccc}
\includegraphics[width=75 mm]{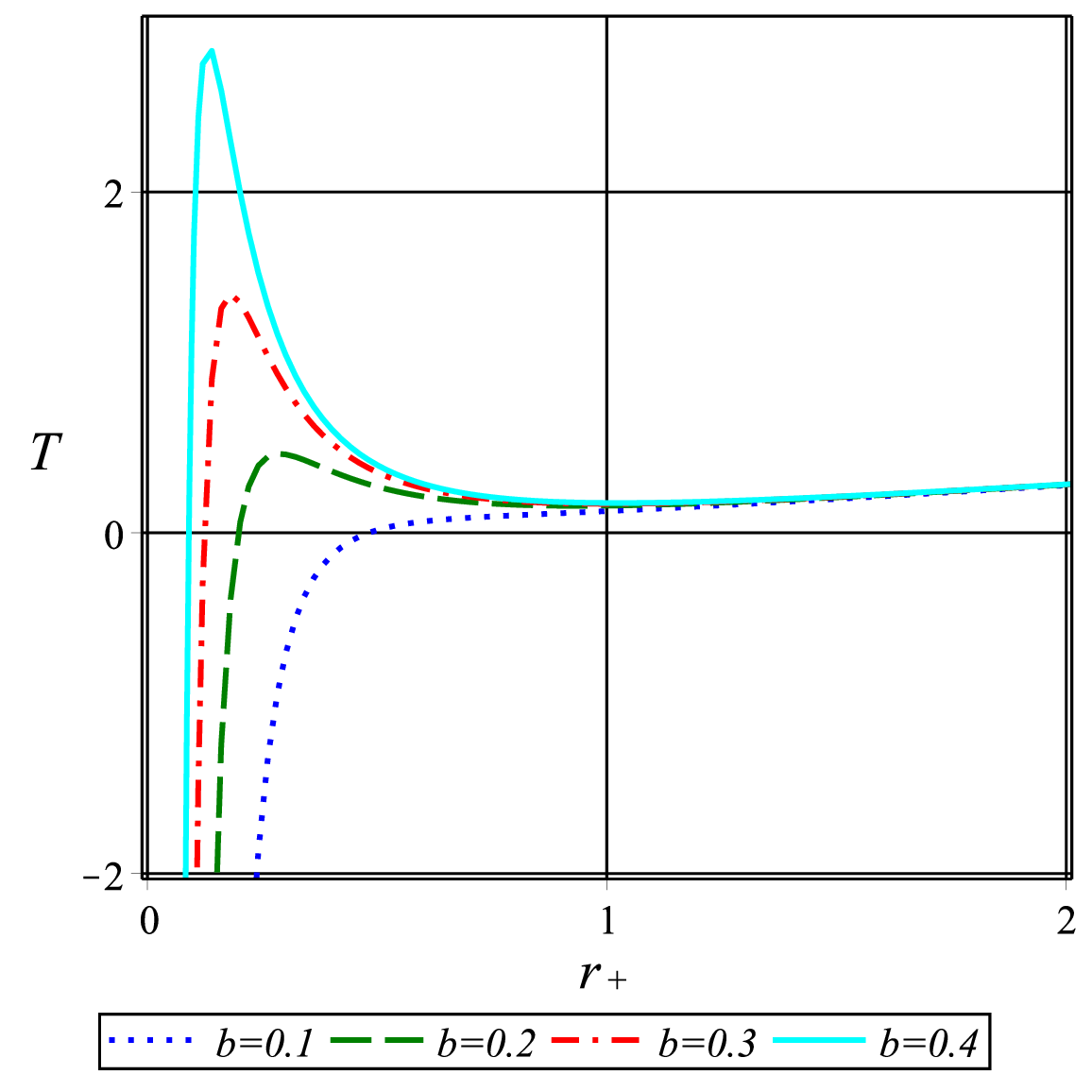}
 \end{array}$
 \end{center}
\caption{Temperature of the charged accelerating AdS black hole in $f(R)$ gravity for different values of $b$ with $a=1$, $q = 0.2$, $R_{0}=-12$, $m = 0.6$.}
 \label{fig2}
\end{figure}

The position of the black hole event horizon is determined  by $f(r) = 0$ (which is illustrated by blue dashed line of the Fig. \ref{fig1}). Also, the parameters $ q $ and $ m $ are related to the black hole electric charge $Q$ and ADM mass $M$  respectively. So, we have the following expressions,
\begin{eqnarray}\label{12}
M &=& \frac{b}{2 K} \left(  r +  \frac{q^{2}}{b r} - \frac{R_{0} r^{3}}{12 (1 - A^{2} r^{2})}\right), \nonumber\\
Q&=&\frac{q}{\sqrt{b } K},\nonumber\\
G_{eff} &=& \frac{G}{1 + f^{\prime}(R)}, \nonumber\\
\Phi&=&\frac{\sqrt{b} q}{r_{h}}.
\end{eqnarray}
We see that  the $f(R)$ gravity give some modifications to mass $M$ and electric charge $Q$ and consequently the corresponding thermodynamic potentials.\\
By considering the Ricci scalar curvature as $R_{0} = - \frac{12}{\ell^{2}} = - 4 \Lambda$, one can find equation (\ref{11}) asymptotically AdS space-time. As it has been discussed in the Ref. \cite{51}, the temperature $T$ will be infinity when the horizon radius $r_{+}\rightarrow \frac{1}{A}$. To avoid this singularity, one can consider the value of $Ar_{+}$ as a constant for simplicity $ A r_{+}=a$. Therefore, we obtain the following  Hawking temperature,
\begin{equation}\label{13}
T= \frac{1}{4 \pi} \left( \frac{d f(r)}{dr} \right) _{r = r_{+}} =\frac{(12m(1+a^{2})-12a^{2}r_{+}-R_{0}r_{+}^{3})br_{+}-12q^{2}}{24\pi b r_{+}^{3}},
\end{equation}
%T= \frac{1}{4 \pi b  r_{+}^{3} ( 1- a^{2} ) } \left[ ( 1- a^{2})^{2} \left( b r^{2}_{+}  - q^{2}  \right) - \frac{b R_{0}}{12} r^{4 }_{+} ( 3- a^{2})  \right],
where $r_{+}$ is largest root of $f(r)=0$ (for example, by selected values of Fig. \ref{fig1} $r_{+}\approx0.65$ corresponding to blue dashed line).\\
In the fig. \ref{fig2} we draw temperature in terms of horizon radius to see effect of parameter $b$. We can see the minimum of the horizon radius blow that the temperature is negative which sounds instability and may interpreted as a non-physical state. Also, we can see a maximum in the temperature which increased its value by increasing $b$. For the large $r_{+}$, there is no $f(R)$ gravity effects as discussed above. Moreover, it is clear that by increasing $b$  the positive range of temperature (physical state) increases too.\\

So, the entropy of the black hole is given by,
\begin{equation}\label{14}
S=\frac{\pi r_{+}^{2} b }{ K ( 1-  a^{2})},
\end{equation}
which is extension of the entropy (\ref{8}). In that case, in order to verify the first law of thermodynamics, one should account the cosmological constant as a dual variable of the thermodynamic pressure \cite{39,42}. In another hand, in the extended phase space,  we will consider the curvature $R_{0}$ as a thermodynamic pressure,
\begin{equation}\label{15}
P= - \frac{ b R_{0}}{32\pi} = \frac{ b \Lambda}{8\pi}=\frac{3 b}{8\pi\ell^{2}}
\end{equation}
and the black hole thermodynamic volume is expressed as,
\begin{equation}\label{16}
V=\frac{4\pi }{3 K }\frac{r_{+}^{3}}{(1- a^{2})}.
\end{equation}
The above quantities are satisfied by the following Smarr formula,
\begin{equation}\label{17}
M = 2 T S + \Phi Q - 2 P V.
\end{equation}
It is consistent with the usual charged accelerating AdS black hole without $f(R)$ correction. Also, the
differential form of the first law of thermodynamics will be as,
\begin{equation}\label{18}
d (\frac{M}{b}) =  T d(\frac{S}{b}) + (\frac{\Phi}{b}) d Q +  V d(\frac{P}{b}),
\end{equation}
Using the Hawking temperature (\ref{13}) and the expression of the pressure (\ref{15}) of this  black
hole, one can derive the equation of state $P(T,  r_{+} )$ as,
\begin{equation}\label{19}
P= \frac{3 b T( 1- a^{2})}{ 2 ( 3 - a^{2}) r_{+}}  -  \frac{3D}{8 \pi  r_{+}^{2}}\left(b +  \frac{q^{2}}{  r_{+}^{2}}\right),
\end{equation}
where $ D \equiv  \frac{( 1- a^{2})^{2}}{( 3- a^{2})} $.\\
From the Equations (\ref{15}) and (\ref{18}), we see that $ R_{0}$ and $ f^{\prime}{(R_{0})}$ give a correction to thermodynamic pressure $P$ of the system as well as the differential form of the first law of thermodynamics. Thus, it is expected that the $f(R)$ modification impress the thermodynamical phase transition of the charged accelerating AdS black hole. In that case, we can see some new features for the thermodynamics of black holes.\\
In the next section,  the $ P - V $ criticality and the phase transitions of such black hole with $f(R)$ gravity are investigated in the extended phase space.
%%%%%%%%%%%%%%%%%%%%%%%%%%%%%%%%%%%%%%%%%%%%%%%%%%%%%%%%%%%%%%%%%%%%%%%%%%%%%%%%%%%%%%%%%%%%%%%%%%%%%%%%%%%%%%%%%%%%%%%%%%%%%%%%%%%%%%%%%%%%%%%%%
\section{Thermodynamical structure in the extended phase space}
In this section, we investigate the thermodynamical structure in the extended phase space of the charged accelerating AdS black hole in the $f(R)$ gravity. Firstly, we discuss the extended phase space and obtain the $ P - V $ criticality of such a black hole. Then, the points of the phase transition related to this black hole are calculated and then, the stability of the system is discussed. Also, the key quantities related
to the corresponding black hole are obtained.
%%%%%%%%%%%%%%%%%%%%%%%%%%%%%%%%%%%%%%%%%%%%%%%%%%%%%%%%%%%%%%%%%%%%%%%%%%%%%%%%%%%%%%%%%%%%%%%%%%%%%%%%%%%%%%%%%%%%%%%%%%%%%%%%%%%%%%%%%%%%%%%%%
\subsection{$P - V$ criticality}
\begin{figure}[h!]
 \begin{center}$
 \begin{array}{cccc}
\includegraphics[width=75 mm]{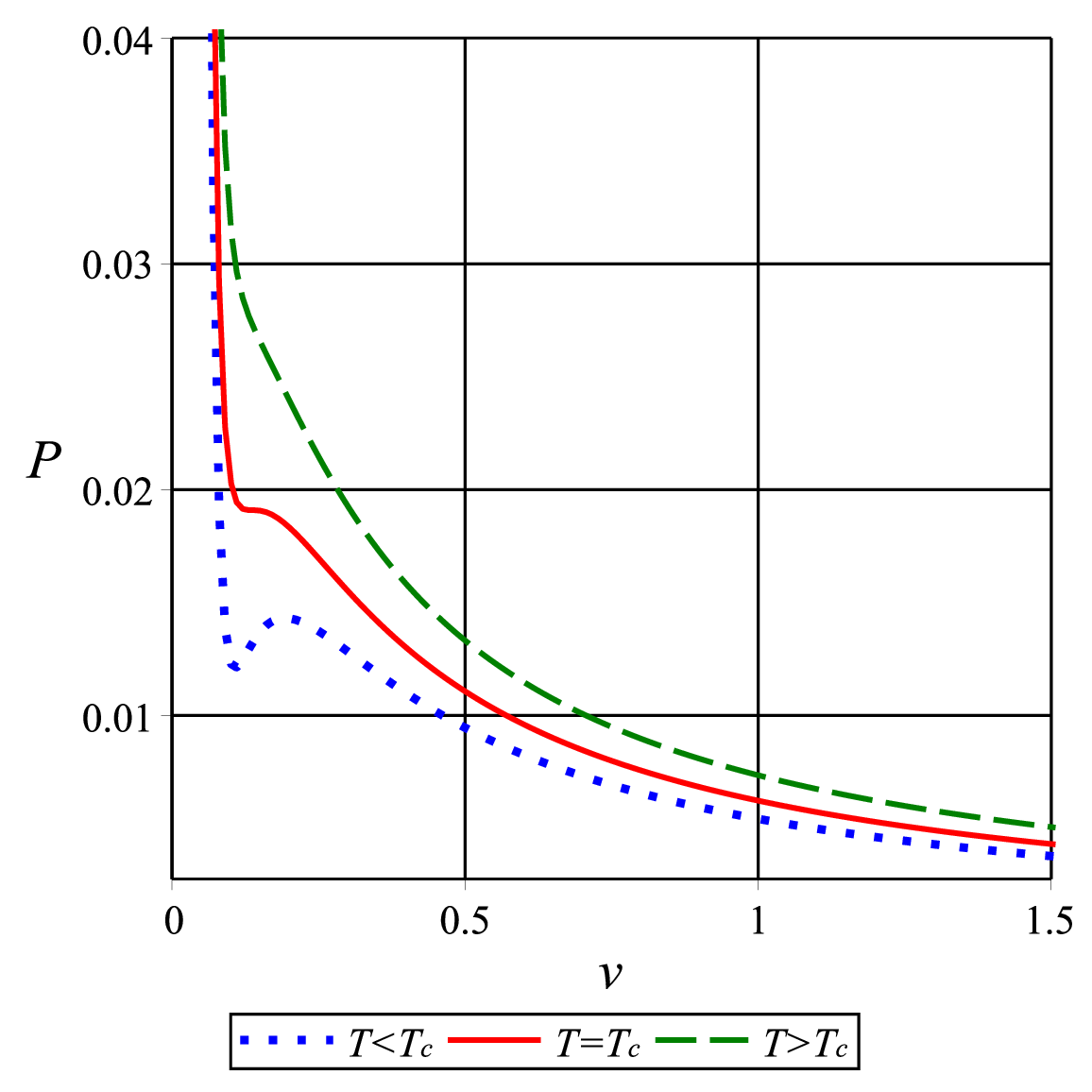}
 \end{array}$
 \end{center}
\caption{Pressure of the charged accelerating AdS black hole with the $f(R)$ gravity for different values of $T$, with $a=0.972$, $b=1$, $q=0.028$ and $T_{c}=0.086$.}
 \label{fig3}
\end{figure}

Now, the $P - V$ criticality and the
phase transition of the charged accelerating AdS black holes are investigated and
 the critical quantities such as $P_{c}$, $T_{c}$ and
$r_{c}$ are calculated. To start, let us turn to review some basic thermodynamic properties of the
charged accelerating AdS black hole.
In the previous section, we obtained the equation of state given by the equation (\ref{19}), which reduced to the following expression in terms of the back hole volume,
\begin{equation}\label{20}
P=    \frac{3 ( 1- a^{2} ) b }{  ( 3 - a^{2}) } \frac{T}{\upsilon}   -    \frac{3 b D}{2 \pi  \upsilon^{2}}  + \frac{6 D q^{2}}{ \pi  \upsilon^{4}},
\end{equation}
where  $\upsilon$  related to the specific volume.\\
The critical points occurs in the inflection point of $P(\upsilon)$ which is determined by the following expression,
\begin{equation}\label{21}
\frac{\partial P}{\partial \upsilon}\bigg|_{ \upsilon = \upsilon_{c}, ~ T=T_{c}}=0~~~;~~~\frac{\partial^{2}
P}{\partial \upsilon^{2}}\bigg|_{\upsilon = \upsilon_{c}, ~ T=T_{c}}=0.
\end{equation}
By using the equations (\ref{20}) and (\ref{21}), we can determine the critical pressure
$P_{c}$, critical specific volume $\upsilon_{c}$  and critical temperature
$T_{c}$,
\begin{equation}\label{22}
\upsilon_{c}= \sqrt{\frac{6}{b}} q, ~~~~~T_{c}= \frac{( 1- a^{2}) }{3 \pi q } \sqrt{\frac{b}{6}}, ~~~~~P_{c}= \frac{b^{2} D}{32 \pi q^{2} }.
\end{equation}
We can obtain a relation between $P_{c},$   $T_{c}$ and $ \upsilon_{c} = B r_{c}$,
\begin{equation}\label{23}
\rho_{c} = \frac{P_{c}\upsilon_{c}}{T_{c}}=\frac{9}{16} \frac{b B ( 1 - a^{2}) ِ}{( 3 - a^{2})}.
\end{equation}
Here, $ \rho_{c}  $ is modified by $f(R)$ gravity. It is obvious that by growing $f^{\prime}(r_{c})$ the critical temperature $T_{c}$ and the pressure $P_{c}$ increase, but the critical volume $ \upsilon_{c} $ decrease.\\
In the case of $q=0$ the van der Waals equation of state recovered where $\frac{3 b D}{2 \pi}$ plays role of strength of the intermolecular interactions in van der Waals gas. However, as illustrated by the Fig. \ref{fig3}, the equation of state (\ref{20}) behaves as van der Waals-like fluid. For the selected values of parameters as Fig. \ref{fig3} we find $T_{c}=0.086$ corresponding to the solid red line of the Fig. \ref{fig3}. The pressure diagram is plotted respect to  the volume with different values of $T$ in Fig. \ref{fig3} when the parameter of the $f(R)$ gravity is constant. Also we find that value of the pressure increased by $b$ when $T>T_{c}$ while decreased by $f(R)$ gravity parameter when $T<T_{c}$.\\

In order to obtain a physical phase transition, all the critical values  $ P_{c} $,  $ T_{c} $ and  $\upsilon_{c}$ must be positive, which lead us to have $ b = 1 + f^{\prime}(R_{0}) > 0, $ which should be satisfied for the mentioned black hole. When $ B \rightarrow  \frac{2 ( 3- a^{2}) }{3 ( 1- a^{2}) b} $, the above relation reduces to the usual relation $\frac{P_{c} \nu_{c}}{T_{c}}=\frac{3}{8}$. This shows that  by introducing charge for the accelerating AdS black hole in $f(R)$ gravity, $\upsilon_{c}$  changes to the form of $ \frac{ ( 3- a^{2}) }{3 ( 1- a^{2}) b} $.
Finally, one can find that the charged accelerating AdS black hole in the $f(R)$ gravity possesses the similar feature of the van der Waals fluid.

%%%%%%%%%%%%%%%%%%%%%%%%%%%%%%%%%%%%%%%%%%%%%%%%%%%%%%%%%%%%%%%%%%%%%%%%%%%%%%%%%%%%%%%%%%%%%%%%%%%%%%%%%%%%%%%%%%%%%%%%%%%%%%%%%%%%%%%%%%%%%%%%%%%%%

\subsection{The stability and phase transition}
This section is devoted to studying the stability conditions for the considered black hole. As an interesting thermodynamical quantity, let us turn to consider the heat capacity. This quantity describes two important features of the black hole, namely the stability and the phase transition.
The heat capacity $C_{Q}$ has the following relation with the entropy and the temperature,
\begin{equation}\label{24}
C_{Q}= T\left( \frac{\partial S}{\partial T}\right)_{Q},
\end{equation}
If  $C_{Q} > 0$,  the black hole is stable, and if $C_{Q} < 0$,  the black hole is in an unstable phase.  By using of the above equation, we can obtain the heat capacity as,
\begin{equation}\label{25}
C_{Q} = \frac{2\pi b r_{+}^{2}( m b A \ell ^{2} (1 +  a ^{2} ) r^{2}_{+}  -   a \ell ^{2} q ^{2}   + a b r ^{4}_{+} )}{K (1-a^{2})( - 2 m b A \ell ^{2}  r^{2}_{+} +  3 a \ell ^{2} q ^{2}   + a b r ^{4}_{+} )}.
\end{equation}
Here, the $ C_{Q} = 0$ corresponds to the first type of the phase transition  which leads to,
\begin{equation}\label{26}
a b r^{4}_{+} +  m b A \ell^{2} (1 + A^{2}r^{2}_{+}) r^{2}_{+}  - a \ell^{2} q^{2} = 0.
\end{equation}
The other type of the phase transition is provided with the divergence points of the heat capacity which is obtained as follows,
\begin{equation}\label{27}
a b r_{+}^{4} - 2 m b A \ell^{2} r^{2}_{+} + 3 a \ell^{2} q^{2} = 0,
\end{equation}
which yields to the following solutions,
\begin{equation}\label{28}
r_{+,1} = \frac{\sqrt{abl(mbAl+\sqrt{m^{2}A^{2}b^{2}l^{2}-3ba^{2}q^{2}})}}{ab},
\end{equation}
and
\begin{equation}\label{29}
r_{+,2} =\frac{\sqrt{abl(mbAl-\sqrt{m^{2}A^{2}b^{2}l^{2}-3ba^{2}q^{2}})}}{ab},
\end{equation}

\begin{figure}[h!]
 \begin{center}$
 \begin{array}{cccc}
\includegraphics[width=60 mm]{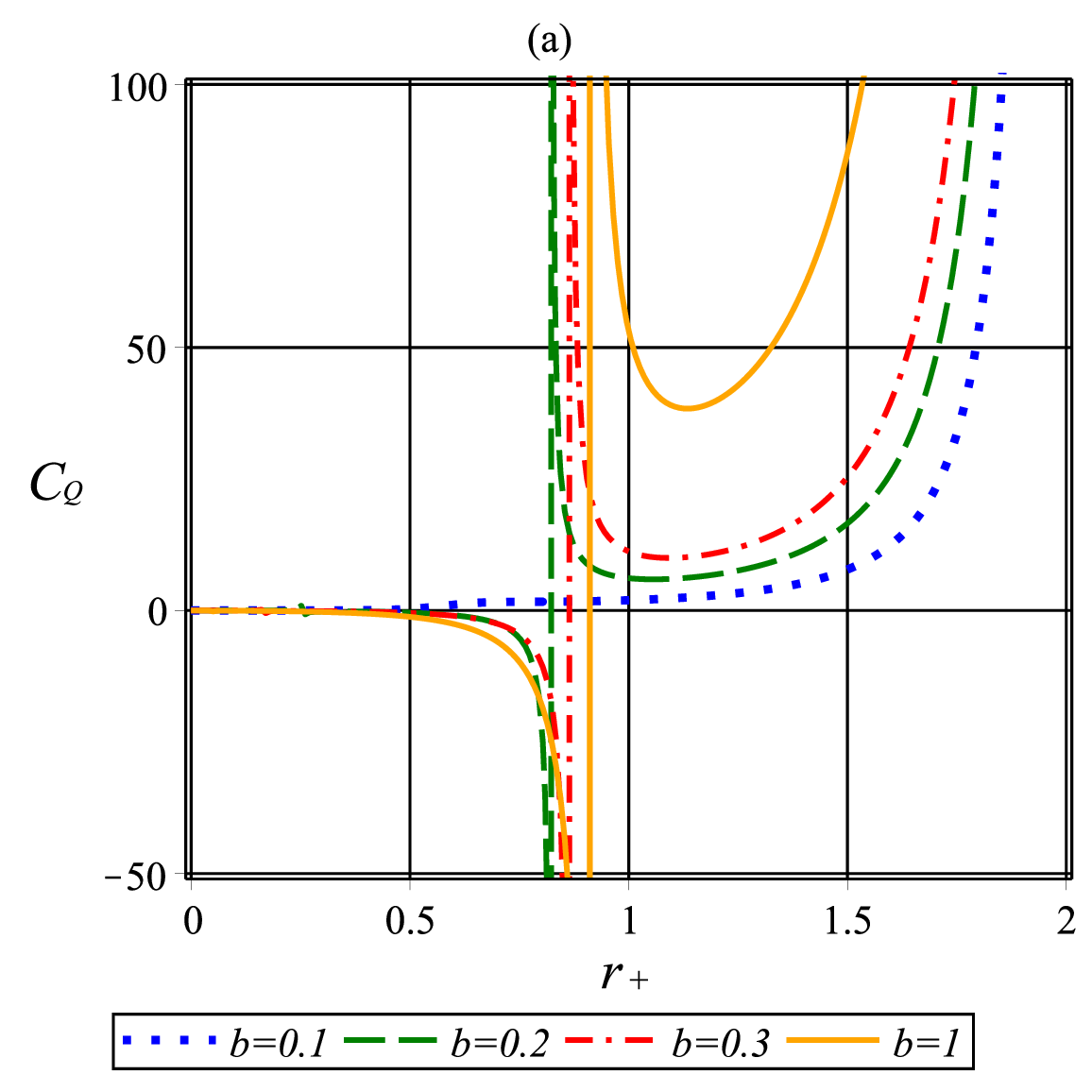}\includegraphics[width=60 mm]{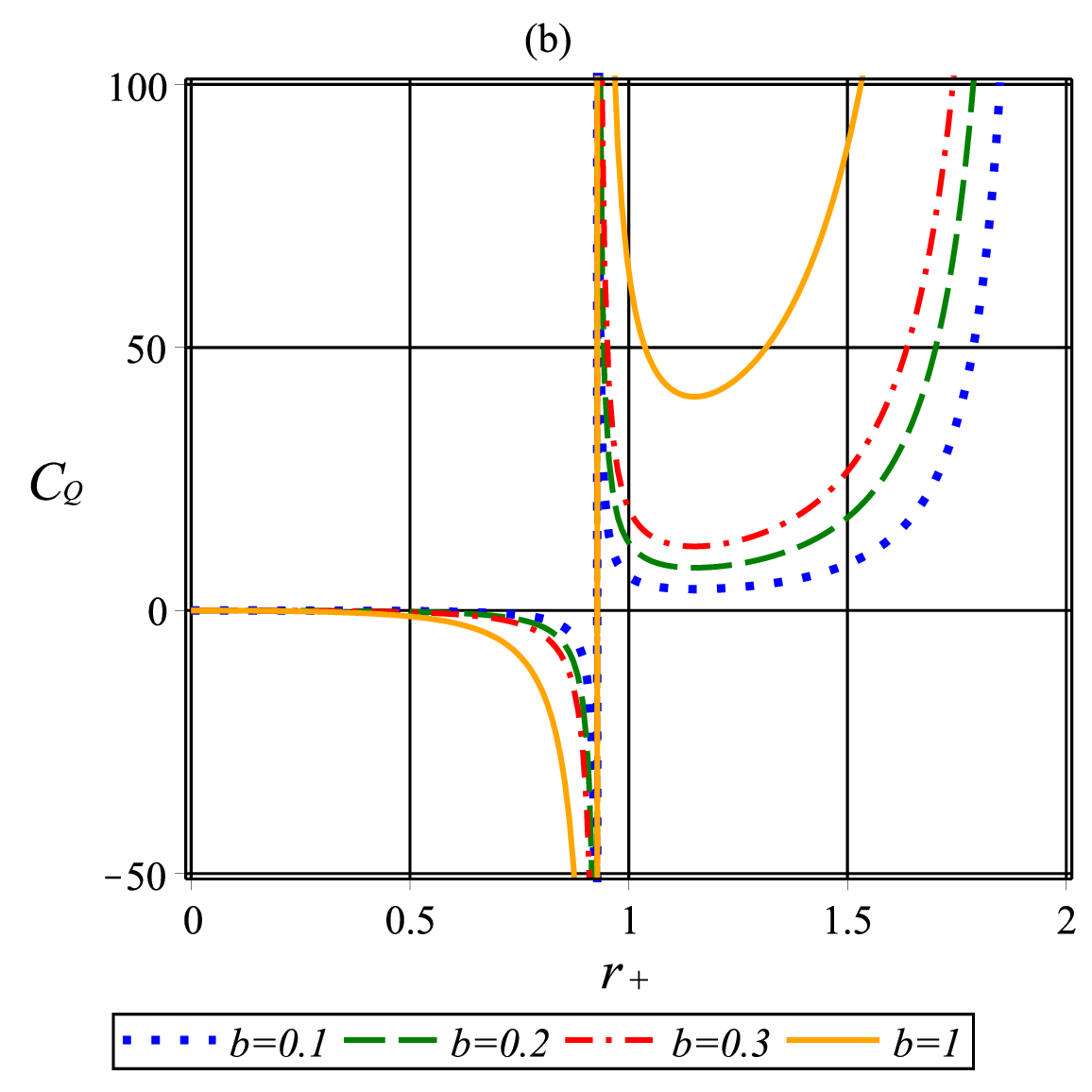}
 \end{array}$
 \end{center}
\caption{The  heat capacity in terms of $r_{+}$ for $R=-12$,  $A=0.5$, and  $m=0.3$. (a)  $ q = 0.1 $  (b) $ q = 0 $.}
 \label{fig4}
\end{figure}

In Fig. \ref{fig4} (a), the behavior of the heat capacity is presented at constant charge for different values of $b$. The plot (b) of Fig. \ref{fig4}  shows uncharged case. In this case we can see phase transition, while there is a situation with small $b$ (dotted blue line of Fig. \ref{fig4} (a)) that the phase transition does not exist.
Here, the effects of $f(R)$ gravity on the stability of this black hole illustrated.
$r_{+,1}$ and $r_{+,2}$ also are presented in both plots of Fig. \ref{fig4} ($r_{+,1}\approx1$ and $r_{+,2}\approx2$ for selected values of parameters).

\section{Joule-Thompson expansion}
In this section, we will examine Joule-Thomson expansion for the charged accelerating AdS black holes of $f(R)$ gravity in the extended phase space.
This expansion is determined by changing the temperature with respect to the pressure, while the enthalpy remains constant during this process. In the AdS space, the black hole mass is interpreted as enthalpy \cite{e1, e2}. Therefore, during the expansion process, the mass of the black hole remains constant. The Joule-Thomson coefficient, which determines the expansion, is given by \cite{52},
\begin{equation}\label{31}
\mu = \left(\frac{\partial T}{\partial P} \right)_{M} = \frac{1}{C_{P}} \left[T \left(\frac{\partial V}{\partial T} \right)_{P} - V \right].
\end{equation}
The cooling and heating regions can be determined by the equation (\ref{31}). The variation of pressure is negative because the pressure during expansion always reduces. The temperature during the process may decrease or increase. Therefore, the temperature determines the sign of $\mu$.  If $\mu$ is positive (negative), cooling (heating) occurs.
The inversion curve, which is obtained at $ \mu = 0 $ for a small infinite pressure, determines the expansion process. Also, it determines the cooling and heating regions  on the $T - P$ screen.\\
Now, we rewrite the black hole equation of state obtained in the previous section, based on the thermodynamic volume. Therefore,  the equation (\ref{13}) becomes,
\begin{equation}\label{32}
T =  \frac{1 }{3 V} \left[ (1-a^{2}) \left(   b \left(  \frac{3  (1-a^{2}) }{4 \pi } \right)  ^{\frac{2}{3}}  V^{\frac{2}{3}}  -  q^{2} \right)  +  2 P \left(   \frac{3 (1-a^{2}) }{ 4 \pi b}\right) ^{\frac{1}{3}} (3 - a^{2}) V^{\frac{4}{3}} \right],
\end{equation}
Using the equation (\ref{32}) and the second side of the equation (\ref{31}), we can obtain the temperature corresponding to a vanishing Joule-Thomson coefficient. Repeated inversion temperature $T_{i}$ one can obtain,
\begin{eqnarray}\label{33}
T_{i} &=&  \frac{1 }{9 V} \left[ (1-a^{2}) \left( - b \left(  \frac{3  (1-a^{2}) }{4 \pi } \right)  ^{\frac{2}{3}}  V^{\frac{2}{3}}  + q^{2} \right)  +  2 P \left(   \frac{3 b ^{2} (1-a^{2}) }{ 4 \pi }\right) ^{\frac{1}{3}} (3 - a^{2}) V^{\frac{4}{3}} \right]\nonumber\\
&=&\frac{1}{12  \pi b r_{+}^{3}  } \left[ (1-a^{2})^{2} \left( -  b r^{2}_{+} +  q^{2} \right)  + \frac{8 \pi }{3} P r_{+}^{4} (3 - a^{2}) \right ].
\end{eqnarray}
Also, the equation (\ref{32}) can be rewritten as follow,
\begin{equation}\label{34}
T = \frac{1}{4  \pi b r_{+}^{3}  } \left[ (1-a^{2})^{2} \left(   b r^{2}_{+} -  q^{2} \right)  + \frac{8 \pi }{3} P r_{+}^{4} (3 - a^{2}) \right ].
\end{equation}
Comparing equations (\ref{33}) and (\ref{34}), leads to the following polynomial equation,
\begin{equation}\label{35}
4 \pi P_{i}  r_{+}^{4}  + 3 D b r^{2}_{+} - 3 D q^{2}  =0.
\end{equation}
Here,  we only consider the real positive root when the inversion pressure $P_{i}$  is zero,
\begin{equation}\label{36}
r_{+,i} =  \frac{\sqrt{ \frac{\sqrt{3 D ( 3 D b + 16 \pi  P_{i} q^{2} ) }}{\pi P_{i} } - \frac{ 3 D b}{\pi P_{i}  }}}{2 \sqrt{2}}.
\end{equation}
By inserting the equation (\ref{36}) in the relation (\ref{33}), the inversion temperature is obtained as,                                                                                                \begin{equation}\label{37}
T_{i} = \frac{\sqrt{P_{i}} D}{\sqrt{2 \pi} b ( 3- a^{2})} \left(  \frac{3 D b^{2} + 8 \pi P_{i} q^{2} - b \sqrt{3 D(3 D b^{2} + 16 \pi P_{i} q^{2} )}}{(\sqrt{3 D(3 Db^{2} + 16 \pi P_{i} q^{2} )} - 3 Db)^{\frac{3}{2}}}\right)
\end{equation}
Then, when the inversion pressure $P_{i}$ is zero, the inversion temperature reaches its minimum:
\begin{equation}\label{38}
T_{i}^{min} = \frac{\sqrt{ 3 b} }{ 4 \pi q} \frac{D^{\frac{3}{2}} (3 D -1) }{( 3- a^{2})}
\end{equation}
By using the equation (\ref{22}), the ratio of  inversion temperature in terms of the critical temperature is given by,
\begin{equation}\label{39}
\frac{T_{i}^{min}}{T_{C}} = \frac{1}{2} \frac{9 \sqrt{2 D} (3 D -1)}{2 ( 3- a^{2})^{2} },
\end{equation}
In the case of $a = 1.12$ we have,
\begin{equation}\label{40}
\frac{T_{i}^{min}}{T_{C}} = \frac{1}{2}.
\end{equation}
The above relation has a perfect agreement with the result of \cite{1706}.\\

\begin{figure}[h!]
 \begin{center}$
 \begin{array}{cccc}
\includegraphics[width=60 mm]{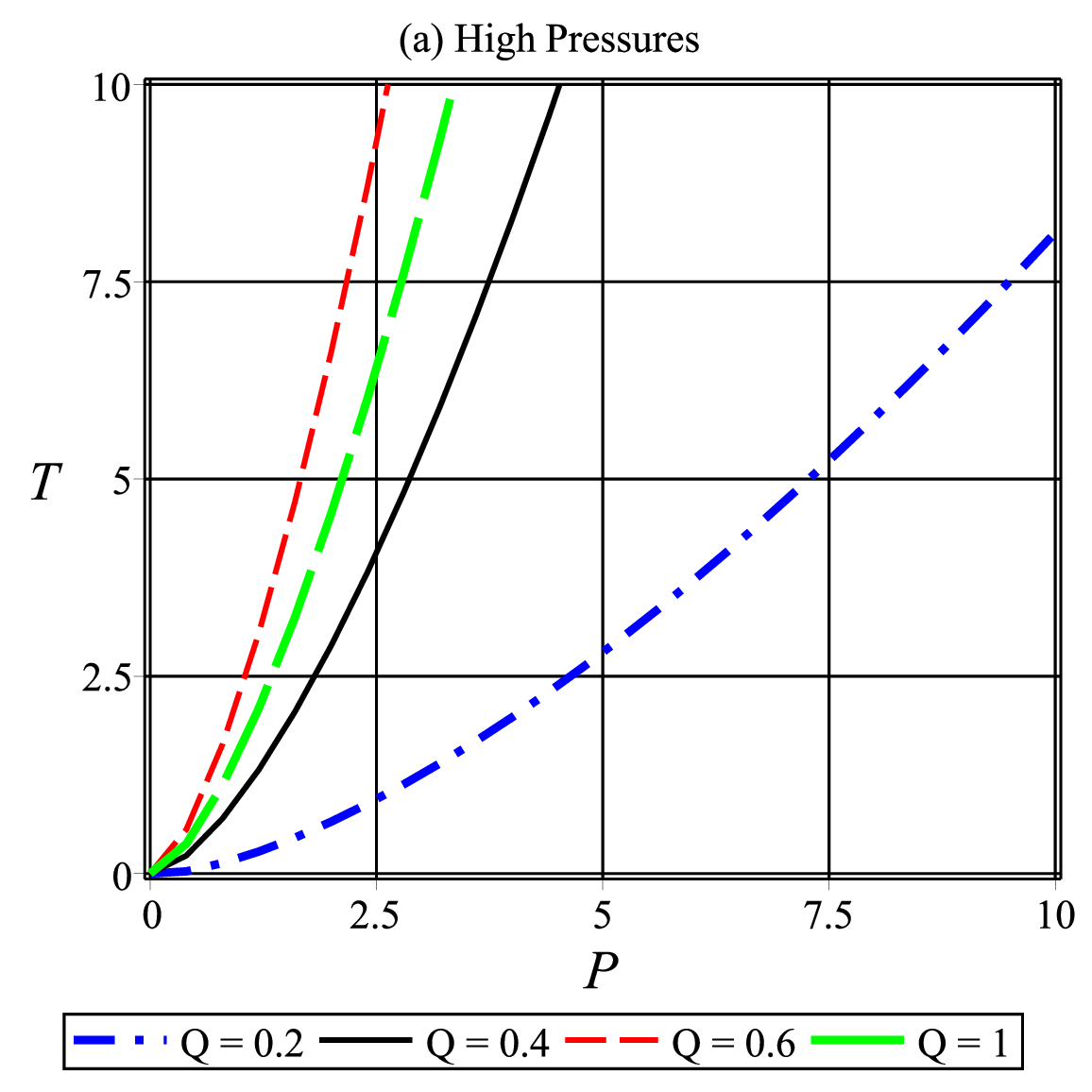}\includegraphics[width=60 mm]{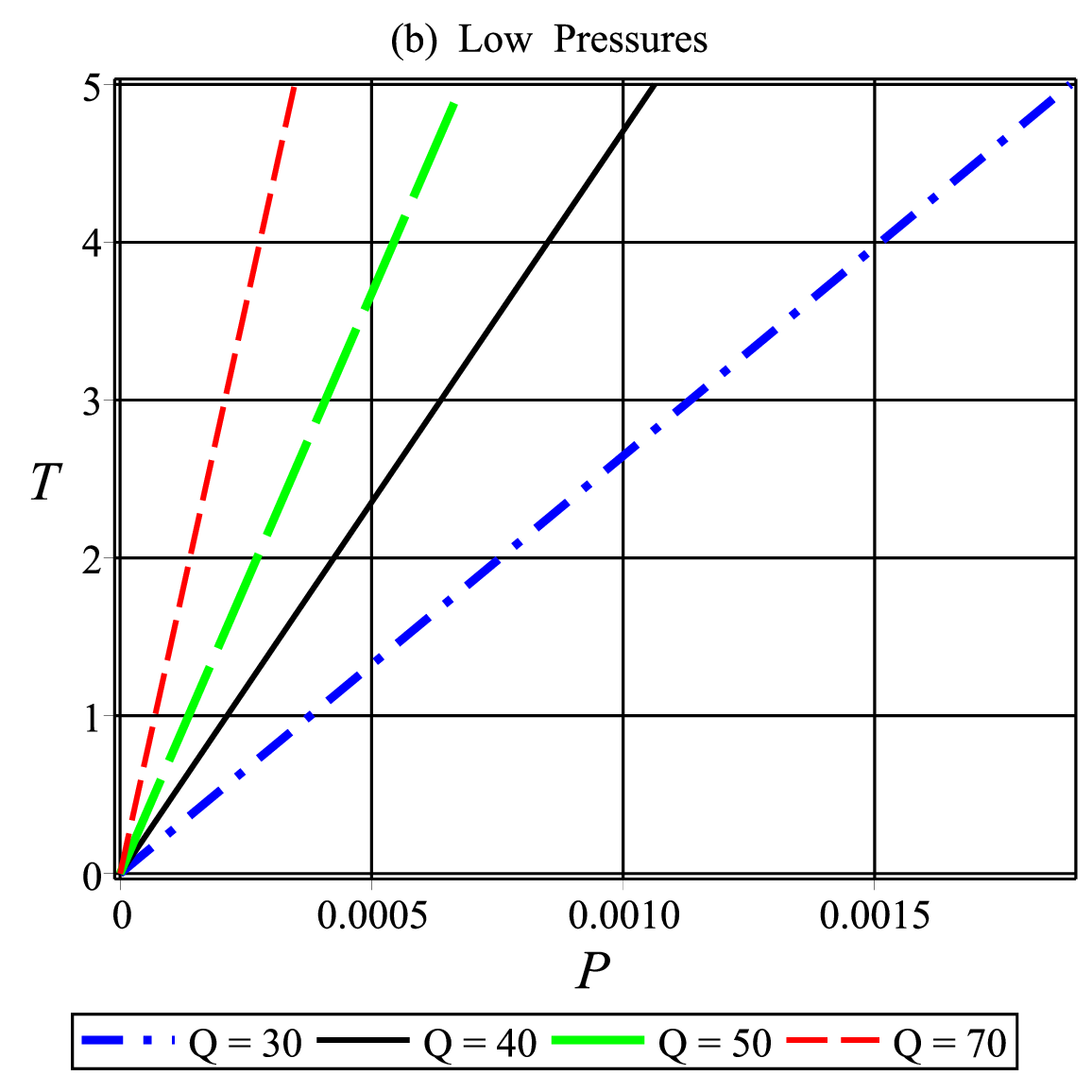}
 \end{array}$
 \end{center}
\caption{The  inversion curves for charged accelerating AdS black hole in the $f(R)$ gravity  with $a = 1.12$, $ b = 1$ and different charges.}
 \label{fig5}
\end{figure}

\begin{figure}[h!]
 \begin{center}$
 \begin{array}{cccc}
\includegraphics[width=60 mm]{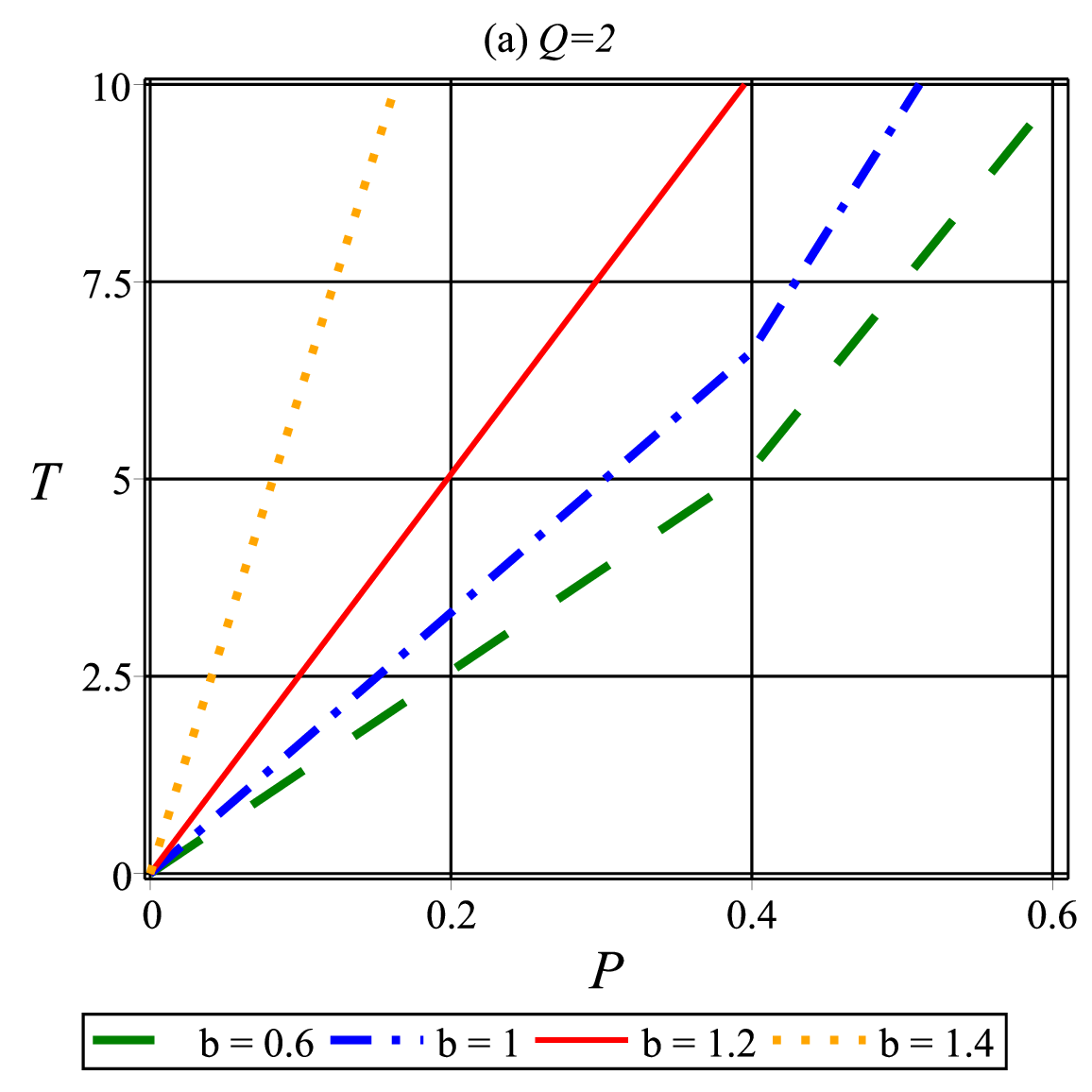}\includegraphics[width=60 mm]{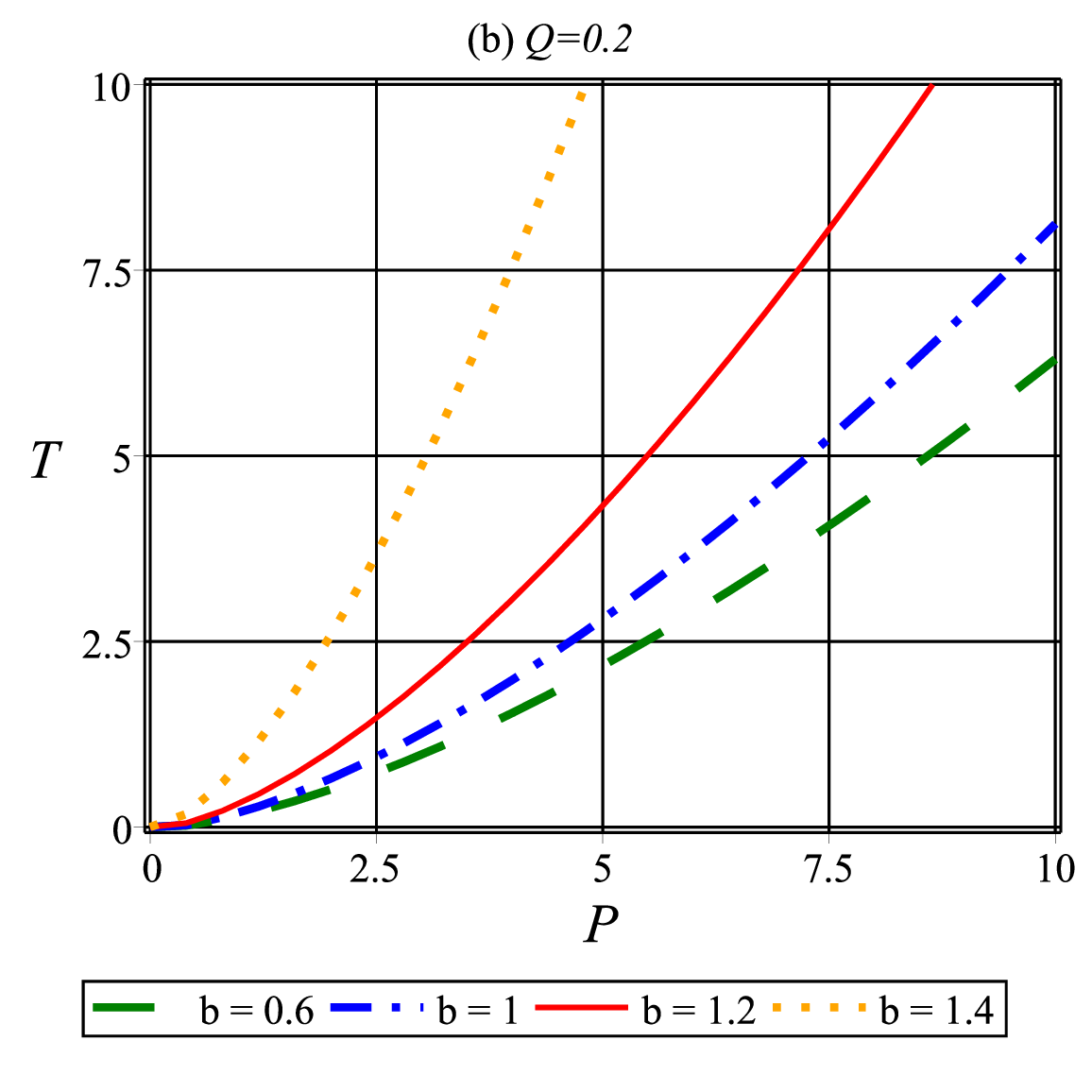}
 \end{array}$
 \end{center}
\caption{The  inversion curves for charged accelerating AdS black hole in the $f(R)$ gravity  with $ a = 1.12 $,  and fixed charges  but  different values of $ b $.}
 \label{fig6}
\end{figure}

In Fig. \ref{fig5}, the inversion curves are plotted for different amounts of the black hole charges at high and low pressures. According to the figure, at low pressure, the inversion temperature increases with constant slope.
Since the inversion temperature is an uniform increasing function, there is no minimum inversion temperature for this black hole. This behavior is fundamentally different from the thermodynamics of the real gases such as the van der Waals type \cite{23}.
Also, the heating and cooling regions  lies above and below these curves, respectively. Although inversion curves have  a similar behavior at the low pressures, they behave differently at high pressures. Also, by  increasing the charge  the temperature increases.\\
In Fig. \ref{fig6}, the inversion temperature is presented for different values of the parameter $ b $ for the fixed charge.  It can be seen that by increasing the parameter $ b $ the inversion temperature increases. However, as $ b $ increases, the slope of the inversion temperature graph becomes more uniform.\\
The Joule-Thomson expansion is considered  by isenthalpic process, i.e. constant mass. We obtained  the isenthalpic curves for different values of black hole mass. The isenthalpic curves  in the $T-P$  plane can be obtained by inserting the event horizon value in the state equation (\ref{19}).

\begin{figure}[h!]
 \begin{center}$
 \begin{array}{cccc}
\includegraphics[width=50 mm]{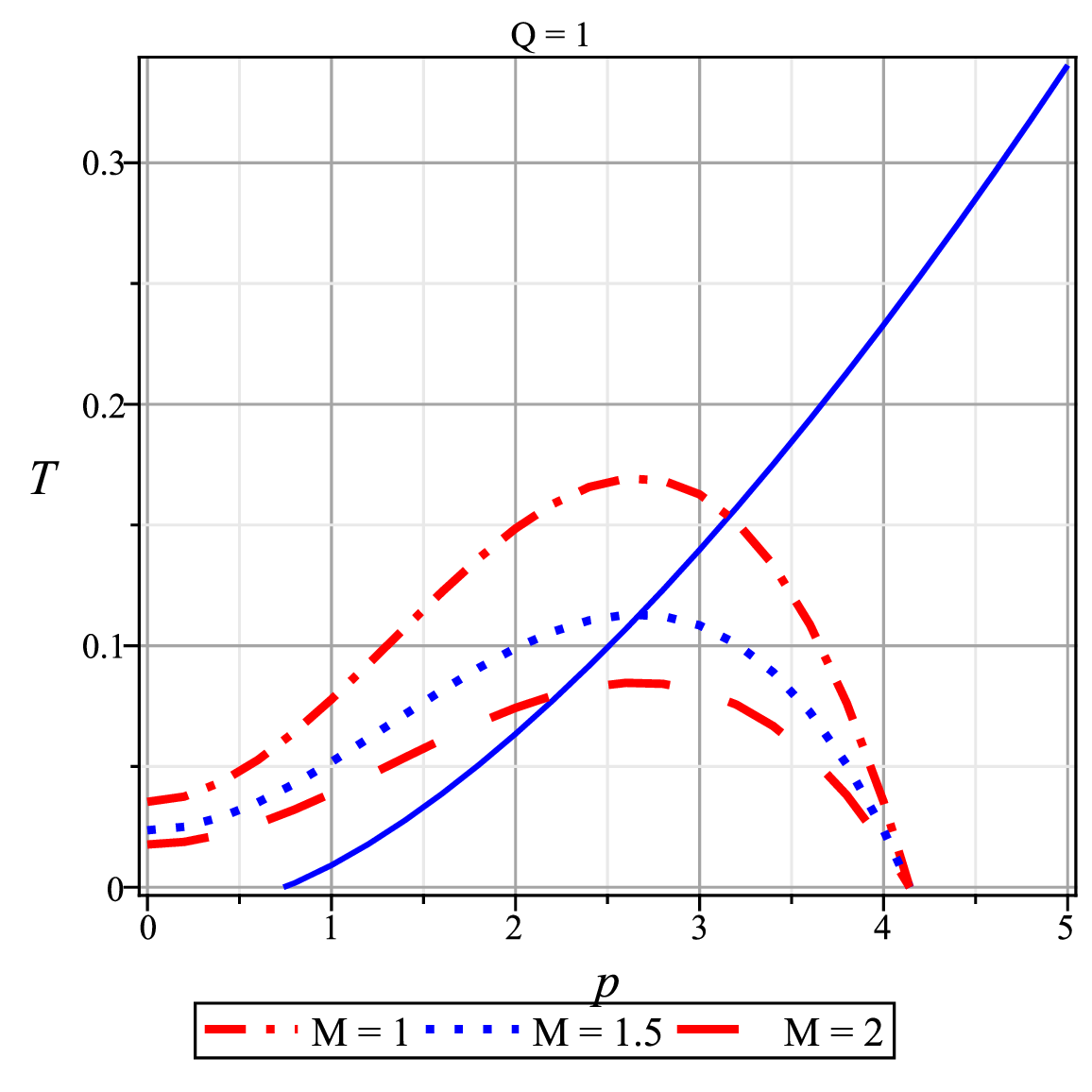}\includegraphics[width=50 mm]{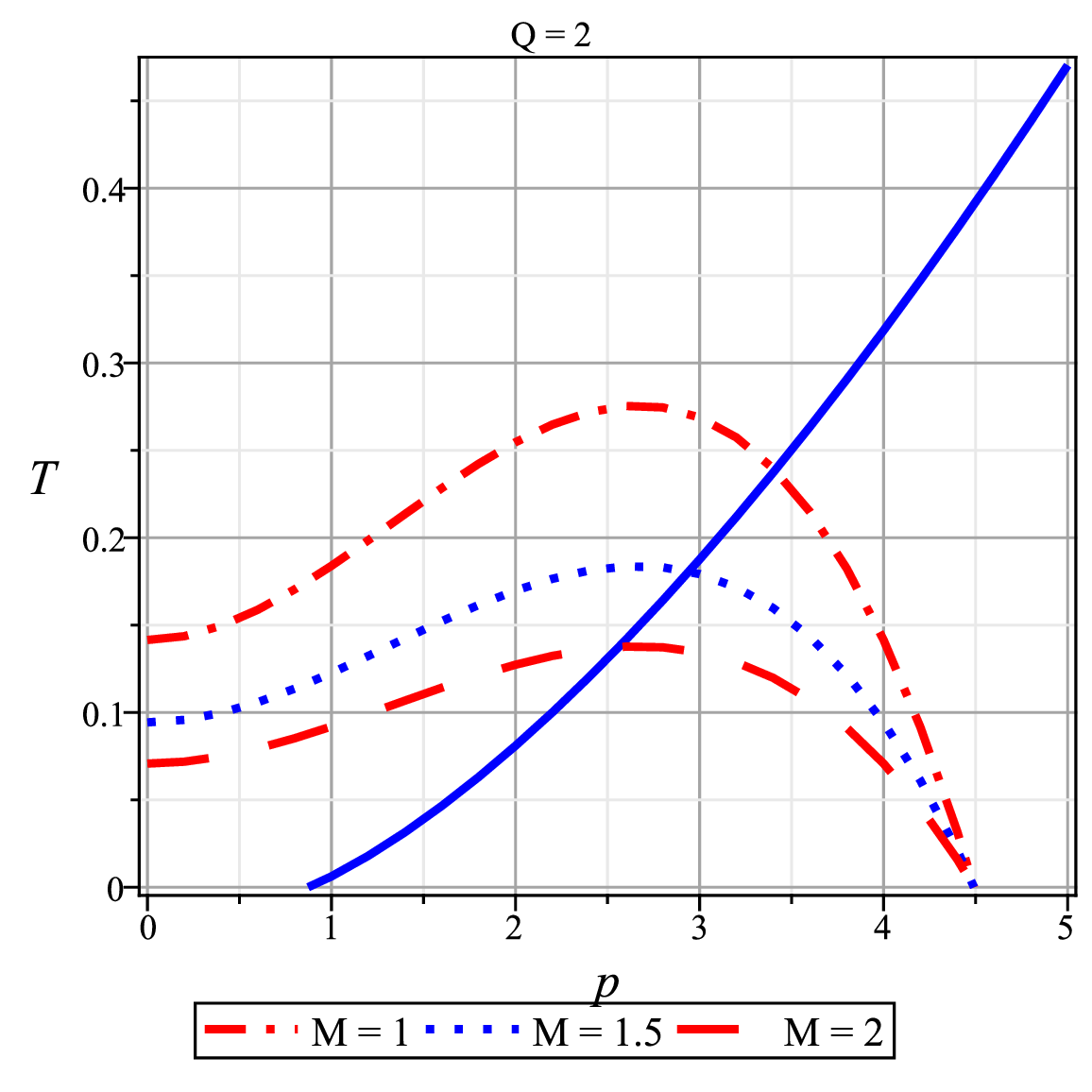}\\
\includegraphics[width=50 mm]{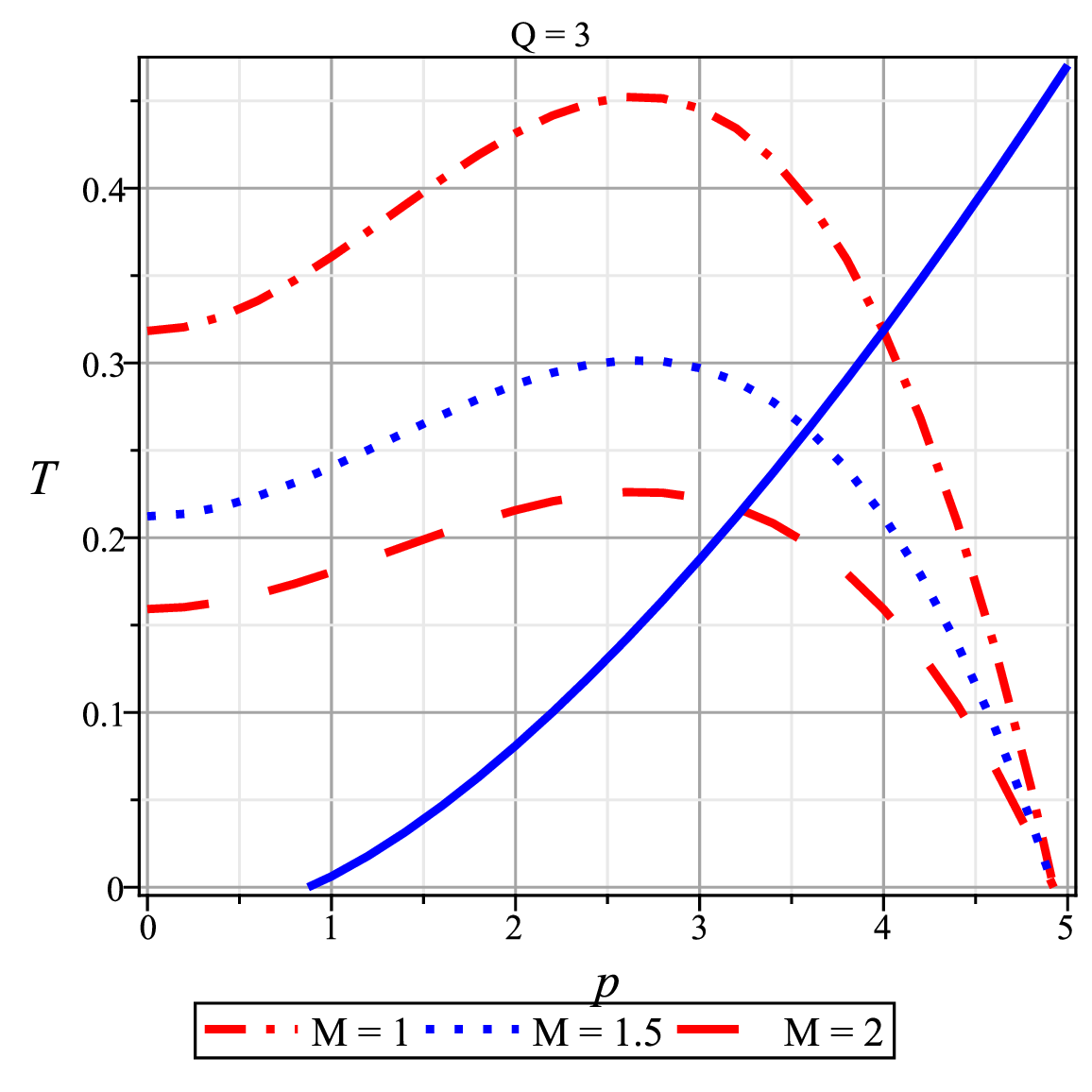}\includegraphics[width=50 mm]{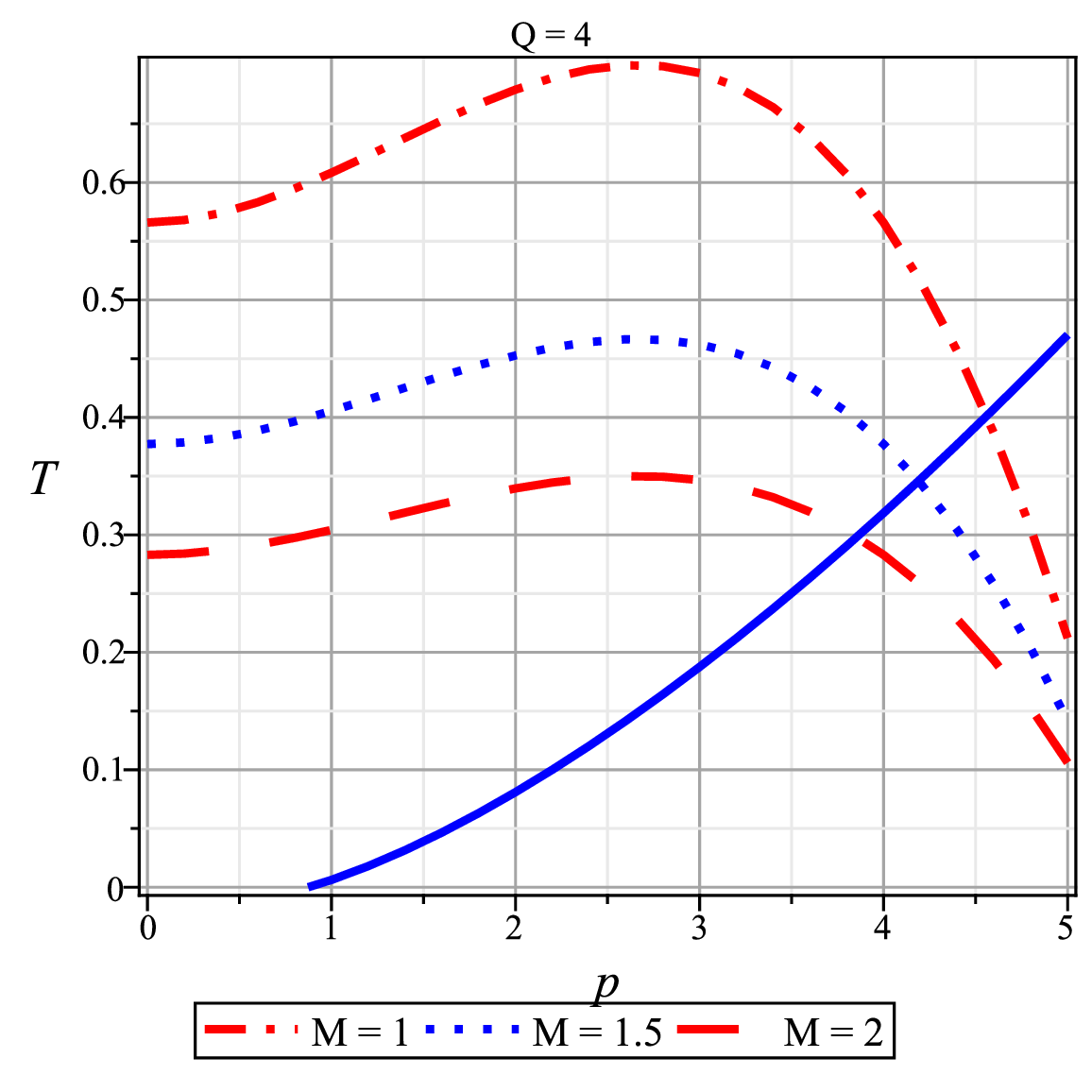}
 \end{array}$
 \end{center}
\caption{The isenthalpic curves for  different values of  black hole mass and  inversion curve ( blue line ) for the corresponding charge. $a = 0.4$, $b=1.6$.}
 \label{fig7}
\end{figure}

Fig. \ref{fig7} shows an example of the isenthalpic curves and the corresponding inversion curves for different charges in the $T-P$ plane. Each diagram contains three curves of different masses and an inversion temperature curve that intersect them at their maximum points. According to the plots, as the charge increases, the inversion curve cut the isenthalpic curve at lower pressures and vice versa is correct for the mass.
The intersection points between the inversion curve and the isenthalpic curves show the reverse points at which the cooling-heating transition point occurs.
That is when the isenthalpic curve crosses the inversion curves, their slop change. Isenthalpic curves have positive slop inside the inversion curves, otherwise, their slop is negative. As a result, the Joule-Thomson coefficient is positive within the inversion curves and cooling occurs in this region.
Therefore, as the pressure increases, the black hole gets warmer.  If the collision point of the curves occurs at higher pressures as the load increases, the black hole warms up later.

%%%%%%%%%%%%%%%%%%%%%%%%%%%%%%%%%%%%%%%%%%%%%%%%%%%%%%%%%%%%%%%%%%%%%%%%%%%%%%%%%%%%%%%%%%%%%%%%%%%%%%%%%%%%%%%%%%%%%%%%%%%%%%%%%%%%%%%%%%%%%%%%%%%%%%%%

\section{Conclusion}
In this paper, the thermodynamics of a charged accelerating AdS black holes have been studied in the $f(R)$ gravity while the Ricci scalar curvature is considered to be constant. The effects of the key parameters of the $f(R)$ gravity on the thermodynamical quantities have been discussed.
By plotting the modified  $f(R)$  gravity as a function of $r$, it has been observed that the gravity is affected by the small horizon radius of the black hole. It means that $f(R)$ gravity effect is like quantum corrections which are important for the small black hole.  In another word, the $f(R)$ gravity terms do not more affect the metric function when $r_{+}$ is large.
It has been perceived that the $f(R)$ gravity give some corrections to the mass $M$ and the electric charge $Q$ and consequently to the corresponding thermodynamical potentials.
Also, it has been seen that $r_{0}$ and $ f^{\prime}{(r_0)}$  produce some corrections to the thermodynamical pressure $P$ of the system as well as the differential form of the first law of thermodynamics. Thus, it is expected that the $f(R)$ modifications influence the thermodynamical phase transition of the charged accelerating AdS black hole.
The extended phase space has been considered and the $P-V$ criticality of such black hole analyzed.  The points of the phase transition related to this black hole have been calculated. It has been perceived that the $P-V$ criticality exists for $T\approx T_{c}$ and also the charged accelerating AdS black holes in the $f(R)$ gravity have van der Waals-like behavior.
It has been observed that with increasing of $f^{\prime}(r_{0})$, both of the critical temperature $T_{c}$ and the pressure $P_{c}$ increased, while the critical volume $ \upsilon_{c} $ decreases. In order to obtain the phase transition, all of the critical values  $ P_{c} $,  $ T_{c} $ and  $ \nu_{c}$ must be positive, which established the condition $ b = 1 + f^{\prime}(R_{0}) > 0$,  which should be satisfied by the considered black hole.\\
It has been found that the $f(R)$ gravity modification changes the ratio $\rho_{c}$ at the critical point. In order to observe the effects of the $f(R)$ gravity on the stability of the system, one must obtain the heat capacity.
Also, it has been observed that the large values of $r_+$ have no significant effect on the graphs, and the small values of $r_+$ has to be chosen to see the changes. It has been represented the first and the second phases transitions corresponding to $r_{+,1}$ and $r_{+,2}$.\\
Also, it has been studied the Joule-Thomson expansion for the charged accelerating AdS black holes of the $f(R)$ gravity in the context of the extended phase space, where the cosmological constant is identified with the pressure.
Here, the black hole mass is interpreted as an enthalpy, so we have assumed that the mass does not change during the Joule-Thomson expansion. Our analysis has shown that the inversion curve always corresponds to the lower curve. This means that the black hole always is in cooling regime and above the inversion curve during the expansion. At last, the cooling and the heating regions have been identified for the different values of the parameter $b$ and the black hole mass $M$. The inversion curve has been plotted for different amounts of the  black hole charges at high and low pressures and investigated the behavior of these curves at different pressures. It seems that at low pressure, as the charge increases, the inversion temperature increases with constant slope.
The ratio of the minimum inversion temperature has been derived in terms of  the critical temperature. In that case, for the certain amounts of  $a$,  it has very good  agreement with the results of the Ref. \cite{1706}. It has been shown that by increasing the parameter  $ b $,  the inversion temperature increases. Also, as  $ b $ increases, the slope of the inversion temperature graph becomes more uniform. By  drawing the isenthalpic curve and the inversion curve on the $T-P$  plane, it has been observed their intersection points where the cooling-heating transition point occurs. Beside, it has been observed that as the charge of the black hole increases, the intersection point grows and the cooling zone increases.\\
As discussed above, $f(R)$ gravity behave like quantum corrections. In order to investigate such behavior further, it is interesting to consider thermal fluctuations \cite{fl1,fl2,fl3} and repeat calculations of this paper. This may prepare us some information about quantum gravity.

\end{document}